\documentclass[twocolumn,showpacs,prl,amsmath]{revtex4-1}
\usepackage{graphicx}
\usepackage{color}
\usepackage{amsmath}
\usepackage{amssymb,srcltx}

\newcommand{\beq}{\begin{equation}}
\newcommand{\eeq}{\end{equation}}
\newcommand{\ket} [1] {\vert#1\rangle}
\newcommand{\bra} [1] {\langle#1\vert}

\newcommand{\dd}{\text{d}}

\begin{document}	

\title{EPR-steering: closing the detection loophole with non-maximally entangled states and arbitrary low efficiency.} 

\author{Giuseppe Vallone}
\affiliation{Department of Information Engineering, University of Padova, via Gradenigo 6/B, I-35131, Padova}

\date{\today}


\begin{abstract}
Quantum steering inequalities allow to demonstrate the presence of entanglement between two
parties when one of the two measurement device is not trusted. In this paper we show that
quantum steering can be demonstrated for arbitrary low detection efficiency by using two-qubit non-maximally entangled states.
Our result can have important applications in one-sided device-independent quantum key distribution.
\end{abstract}





\maketitle

{\it Introduction - }
Entanglement is the most peculiar feature of quantum mechanics and its detection represent an
important task in quantum information.
In order to detect entanglement between two parties (called Alice and Bob) it is possible
to use the entanglement witness method \cite{horo96pla,terh00pla,toth05prl,horo09rmp}, allowing to verify the presence of entanglement when both Alice and Bob
devices are know and trusted (and they also known the dimension of the quantum state they share). 
They can measure an entanglement witness operator $W$ and, when its expectation values 
is negative, the shared state $\rho_{AB}$ is entangled and it cannot be written
as $\rho_{AB}=\sum_\lambda p_\lambda\rho^\lambda_A\otimes
\rho^\lambda_B$. Equivalently, for entangled states, the conditional probabilities cannot be written as 
\beq
P(a,b|{\bf A}_k,{\bf B}_j)\!\!=\!\!\sum_\lambda p_\lambda\text{Tr}[\Pi^k_a\rho^\lambda_A]
\text{Tr}[\Pi^j_b\rho^\lambda_B]\,\,\,\,\text{(separable)}
 \eeq
Here we label the Alice and Bob measurements as  ${\bf A}_k$  and ${\bf B}_j$,
while $a$ and $b$ are the corresponding outputs. 
 $P(a,b|{\bf A}_k,{\bf B}_j)$ is the probability of obtaining the outputs $a$ and $b$ when Alice and Bob choose
 the measurements ${\bf A}_k$ and ${\bf B}_j$, while $\Pi^k_a$ (and similarly for $\Pi^j_b$) is the projector
 into the eigenstate of ${\bf A}_k$ with eigenvalue $a$.

On the other side, it is  well known that the violation of a Bell inequality \cite{bell64phy,clau69prl,clau74prd} is equivalent to the 
detection of entanglement between Alice and Bob with untrusted devices.
In this scenario, Alice and Bob don't know how their measuring device work and they don't know what is the state
they share: however, if a particular combination of their measurement outputs violate some Bell inequality
they can prove that the shared state is entangled. If the Bell inequality is violated no 
local hidden variable (LHV) model can explain the correlation. Formally, a LHV model is written as:
\beq\label{Pab}
P(a,b|{\bf A}_k,{\bf B}_j)=\sum_\lambda p_\lambda A_k(\lambda)B_j(\lambda)\quad\text{(LHV model)}
 \eeq
In the rhs of equation \eqref{Pab}
  $\lambda$ is the {\it hidden variable} with probability $p_\lambda$
 and $A_k(\lambda)$ and $B_j(\lambda)$ are the so called response function depending on $\lambda$ and taking values on the
 possible measurement outcomes.
The possibility of revealing entanglement with untrusted measuring device has important consequences for the
 so called device-independent (DI) secure Quantum Key Distribution (QKD) \cite{acin07prl,masa11nco,luca12pra}.
Alice and Bob can establish a secret key even if the shared state and their measuring device 
where provided by an evestropper.

EPR-Steering inequalities lie in between Entanglement witness and Bell inequality: they allows to demonstrate entanglement when only one of
the two measuring device is trusted~\cite{wise07prl}. Steering has attracted a lot of attention in the last years
\cite{jone07pra,walb11prl,smit12nco,witt12njp,benn12prx,hand12nph,chen12qph}.
Let's consider the case of trusted Bob's device.
If a steering inequality is violated, the shared stated cannot be written as a Local Hidden State (LHS) model:
\beq\label{LHS}
P(a,b|{\bf A}_k,{\bf B}_j)\!\!=\!\!\sum_\lambda p_\lambda A_k(\lambda)
\text{Tr}[\Pi^j_b\rho^\lambda_B]\,\,\,\,\text{(LHS model)}
 \eeq
As noticed in~\cite{bran12pra}, steering is also relevant for QKD: 
precisely, violating an EPR-steering inequality allow to demonstrate the security in one-sided DI
secure QKD, in which Bob's detection device is trusted while Alice's apparatus is not. 

In order to experimentally violate a Bell or steering inequality, it is crucial to close the so called {\it loopholes}: 
the locality~\cite{weih98prl} and freedom-of-choice~\cite{sche10pnas} loopholes are not important in the framework of cryptography, 
because it is a necessary assumption of security that Alice's and Bob's laboratory have no information leakage.
The most crucial loophole is the so called {\it detection loophole}: due to the low detection efficiency of typical
two photon experiments, the inequality is calculated by using 
the additional assumption of fair sampling. Without fair sampling,
at least 83\% efficiency is required to violate the CHSH inequality~\cite{clau69prl} with maximally entangled state,
while for a large class of two-party Bell inequalities the threshold detection efficiency can be lowered
by using non maximally entangled state~\cite{eber93pra,vall11qph}.

In this paper we show that a steering inequality equivalent to the one introduced in~\cite{cava09pra,saun10nph} and 
experimentally violated by using the fair-sampling assumption in \cite{saun10nph} can be violated
with arbitrary detection efficiency by using non-maximally entangled states (NMES).
Note that in~\cite{witt12njp} a loophole-free steering was demonstrated by using an inequality requiring at least 33\% efficiency,
while arbitrary loss tolerant inequality were proposed (and violated) in~\cite{benn12prx}: however, the latter inequalities require
that Alice declare when she detect a photon (or equivalenty when she can "steer" Bob's state) and
cannot be applied when Alice's device is not allowed to give null result.

{\it Rewriting the Steering inequality - }Let's consider the particular case in which Bob subsystem is a qubit and the Alice measurement devices have
two outputs, namely $+1$ and $-1$. Alice and Bob can respectively 
choose between $n$ different measurements ${\bf A}_k$ and $\sigma_{b_k}$,
where $\sigma_{b_k}\equiv \vec b_k\vec \sigma$, $\vec\sigma=\{\sigma_1,\sigma_2,\sigma_3\}$ are the Pauli matrices 
and the $\vec b_k$'s are three-dimensional unit length vectors. 
We consider the situation in which the Alice measurement device cannot give null result:
when Alice chooses a measurement the device is answering with $+1$ or $-1$. 
The inequality introduced in \cite{cava09pra,saun10nph} is written as 
\beq
\label{inequ-orig}
S_n=\frac{1}{n}\sum^n_{k=1}\langle {\bf A}_k\otimes\sigma_{b_k}\rangle\leq C_n\,,
\eeq
with $S_n$ the steering parameter.
If the correlation between Alice and Bob can be described by LHS model, the value of $S_n$ is bounded by 
$C_n=\frac{1}{n}\text{max}_{\{A_k\}}\lambda(\sum_k A_k\sigma_{b_k})$, where $\lambda(\hat O)$ is the
maximum eigenvalues of the operator $\hat O$ and $A_k=\pm1$ (see \cite{saun10nph}). 
The corresponding pure state eigenvectors can be used as $\rho^\lambda_B$ in the LHS model to saturate the
bound in~\eqref{inequ-orig}.

\begin{figure}[tbp]
\includegraphics[width=4.2cm]{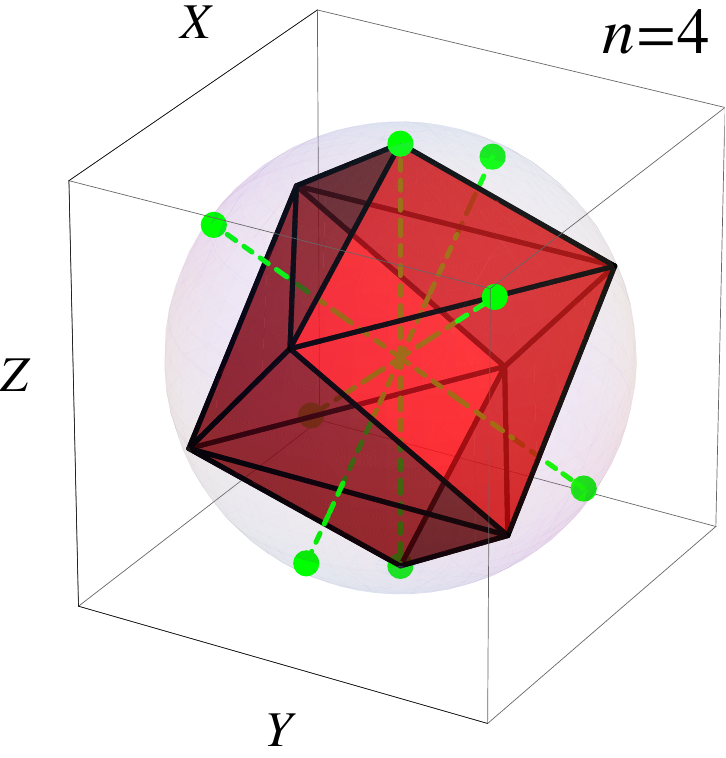}
\includegraphics[width=4.2cm]{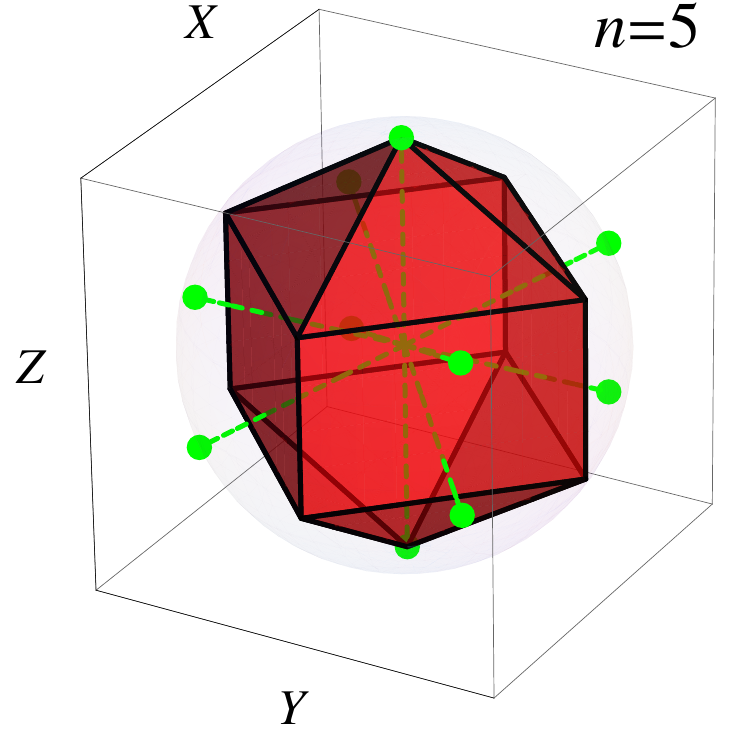}
\caption{Bob measurement are defined by the vertex of the two solid figure for $n=4$ and $n=5$ respectively. Green points
connected with dashed lines to the origin represent the pure states that saturate the LHS bound in \eqref{inequ-orig}.}
\label{fig:measure}
\end{figure}

Note that $C_n$ depends on the choice of observables made by Bob.
For low $n$ values, if the  $\pm\vec b_k$ are chosen as the vertex of platonic solid, 
the square for $n=2$, the octahedron for $n=3$, the
 icosahedron for $n=6$ and the dodecahedron for $n=10$,
the $C_n$ values take the following values~\cite{saun10nph}:
\beq
\begin{aligned}
&C_2=\frac{1}{\sqrt{2}},\quad
C_3=\frac{1}{\sqrt{3}},\quad 
C_6=\frac{1+\sqrt{5}}{6},\quad \\
&C_{10}=\frac{3+\sqrt{5}}{10},\quad \cdots,\quad C_{n\rightarrow\infty}=\frac{1}{2}\,.
\end{aligned}
\eeq 
With $n=4$ measurements it was shown in~\cite{saun10nph} that a bound of $\frac{1}{\sqrt{3}}$
can be achieved  if the  $\vec b_k$ are chosen as the vertex of  a cube.
However, it is possible to find a better choice of the measuring vectors: take $\vec b_1=(0,0,1)$, 
and choose the other three vector as $\vec b_k=(\sin\beta_0\cos\phi_k,\sin\beta_0\sin\phi_k,\cos\beta_0)$ with
$\cos\beta_0=\frac{\sqrt{13}-1}{6}$ and $\phi_2=0$, $\phi_3=2\pi/3$, $\phi_4=4\pi/3$. 
The same strategy can be applied with 5 measurements (with the same $\beta_0$ and 
$\phi_2=0$, $\phi_3=\pi/2$, $\phi_4=\pi$, $\phi_5=3\pi/2$). In figure \ref{fig:measure} we show
the directions of the measurements as the vertices of the solid figures. 
With these settings we obtain:
\beq
C_4=\frac{1+\sqrt{13}}{8}<\frac{1}{\sqrt{3}}\,,
\qquad
C_5=\frac{1+2\sqrt{13}}{15}\,.
\eeq
Note than the $C_n$ series is a decreasing series converging toward $\frac12$.
For any choice of Bob obserables, the inequality \eqref{inequ-orig} can violated by using a two qubit 
maximally entangled singlet state when Alice chooses the measurement
${\bf A}_k=\vec a_k\vec \sigma$ with $\vec a_k=-\vec b_k$: in this case  $S_n=1$.

From the inequality \eqref{inequ-orig}
 we can derive a simpler inequality involving only one output on the Alice and Bob side.
To do so it is sufficient to notice that the single qubit observables can be written as
$\sigma_{b_k}=2\Pi^b_{k}-\openone$ where $\Pi^b_{k}$ is the projection operator on the $+1$ eigenstate of
$\sigma_{b_k}$. Since Alice apparatus alway produces an output, we have 
$P(1,b|{\bf A}_k,{\bf B}_j)-P(-1,b|{\bf A}_k,{\bf B}_j)=2P(1,b|{\bf A}_k,{\bf B}_j)-1$.
Then the correlation term $\langle {\bf A}_k\sigma_{b_k}\rangle$ can be rewritten as
$4P(1,1|{\bf A}_k,\Pi^b_{k})-2P_A(1|{\bf A}_k)-2P_B(1|\Pi^b_{k})+1$. Since only $+1$ outcomes are involved in both
Alice and Bob side,
we simplify the notation as
$\langle {\bf A}_k\sigma_{b_k}\rangle\equiv 4p({\bf A}_k,\Pi^b_{k})-2p({\bf A}_k)-2p(\Pi^b_{k})+1$.
The inequality \eqref{inequ-orig} can be rewritten as
\beq\label{new-ineq}
S'_n=\frac{1}{n}\sum^n_{k=1}\left[2p({\bf A}_k,\Pi^b_{k})-p({\bf A}_k)-p(\Pi^b_{k})\right]\leq C'_n\,,
\eeq
with $C'_n=\frac{C_n-1}{2}$.
The relation between the previous inequality and \eqref{inequ-orig}
is the same that holds between the Clauser-Horne (CH)~\cite{clau74prd} and the CHSH~\cite{clau69prl} inequality:
while in \eqref{inequ-orig} correlations between two-output measurements are involved,
the new inequality \eqref{new-ineq} involves only terms containing $+1$ outputs.
{Since the Bob measuring device is trusted his measurement can be described by a 
well characterized quantum observable and it is possible to consider only the events in which
Bob obtains a non-null result (+1 or -1)~\cite{benn12prx}. 
Moreover, Alice apparatus can be simplified to have only the +1 output (in fact losses are equivalent to
-1 output in the inequality \eqref{new-ineq}).}

Let's now suppose that an honest Alice want to convince Bob about her ability to steer his state by using a two-qubit
entangled state $\rho_{AB}$ with reduced states $\rho_A=\text{Tr}_B[\rho_{AB}]$ and
$\rho_B=\text{Tr}_A[\rho_{AB}]$. Unfortunately Alice has an inefficient measuring device with $\eta$ efficiency.
Alice use the projectors $\Pi^a_k$ as measurement.
In this case $p({\bf A}_k,\Pi^b_{k})=\eta\text{Tr}_{AB}[\Pi^a_k\otimes\Pi^b_k\rho_{AB}]$,
$p({\bf A}_k)=\eta\text{Tr}_A[\Pi^a_k\rho_A]$ and $p(\Pi^b_k)=\text{Tr}_B[\Pi^b_k\rho_B]$. 
Alice is able to demonstrate steering only if her efficiency satisfy $\eta>\eta^{(n)}_c$ with the critical efficiency
given by
\begin{align}
\eta^{(n)}_c&=\frac{C'_n+\frac{1}{n}\sum^n_{k=1}\text{Tr}_B[\Pi^b_k\rho_B]}
{\frac{1}{n}\sum^n_{k=1}\left(2\text{Tr}_{AB}[\Pi^a_k\Pi^b_k\rho_{AB}]-\text{Tr}_A[\Pi^a_k\rho_A]\right)}
\\
\label{etac2}
&=\frac{C_n+\frac{1}{n}\sum^n_{k=1}\langle \vec b_k\vec \sigma\rangle_{\rho_B}}
{\frac{1}{n}\sum^n_{k=1}\left[\langle \vec a_k\vec \sigma\otimes \vec b_k\vec \sigma\rangle_{\rho_{AB}}+
\langle \vec b_k\vec \sigma\rangle_{\rho_B}\right]}\,,
\end{align}
where $\rho_B$ is the reduced state on Bob side.
By using maximally entangled state 
the best critical efficiency is given by
\beq
\eta^{(n)}_c=C_n\qquad\text{(for maximal entangled states)}
\eeq
In fact for maximally entangled state we have $\rho_B=\frac12\openone$ and 
we get $\langle \vec b_k\vec \sigma\rangle_{\rho_B}=0$ $\forall\vec b_k$. Moreover,
by carefully choosing the $\vec a_k$'s it is possible to 
obtain $\langle \vec a_k\vec \sigma\otimes \vec b_k\vec \sigma\rangle_{\rho_{AB}}=1$  $\forall k$
and the best $\eta^{(n)}_c$ is equal to $C_n$.

{\it Reducing $\eta_c$ with NMES -} We now demonstrate that by using non-maximally entangled
 state the critical efficiency can be lowered.
Let's consider the following non-maximally entangled state:
\beq\label{psi}
\ket{\psi}=\cos\frac\theta2\ket{01}-\sin\frac\theta2\ket{10}\,,
\eeq
and define the measuring projector as $\Pi^a_k=\ket{a_k}\bra{a_k}$
and $\Pi^b_k=\ket{b_k}\bra{b_k}$ with
\beq
\begin{aligned}
&\ket{a_k}=\sin\frac{\alpha_k}2\ket{0}-e^{i\varphi_k}\cos\frac{\alpha_k}2\ket{1}\,,
\\
&\ket{b_k}=\cos\frac{\beta_k}2\ket{0}+e^{i\phi_k}\sin\frac{\beta_k}2\ket{1}\,.
\end{aligned}
\eeq
with $0\leq \phi_k\leq2\pi$, $0\leq \varphi_k\leq2\pi$ and $0\leq \alpha_k\leq\pi$. 
The parameter $0\leq \theta\leq \pi/2$ is an entanglement monotone \cite{horo09rmp} and can be related to the 
content of entanglement of the state $\ket{\psi}$.
In equation \eqref{etac2} only the denominator depends on the $a_k$'s. It is maximized (and then $\eta_c$ 
is minimized) when the $a_k$'s
are chosen such that:
\beq\label{alpha}
\tan\alpha_k=\sin\theta\tan\beta_k\,,\qquad\varphi_k=\phi_k\,.
\eeq

\begin{figure}[tbp]
\centering\includegraphics[width=7cm]{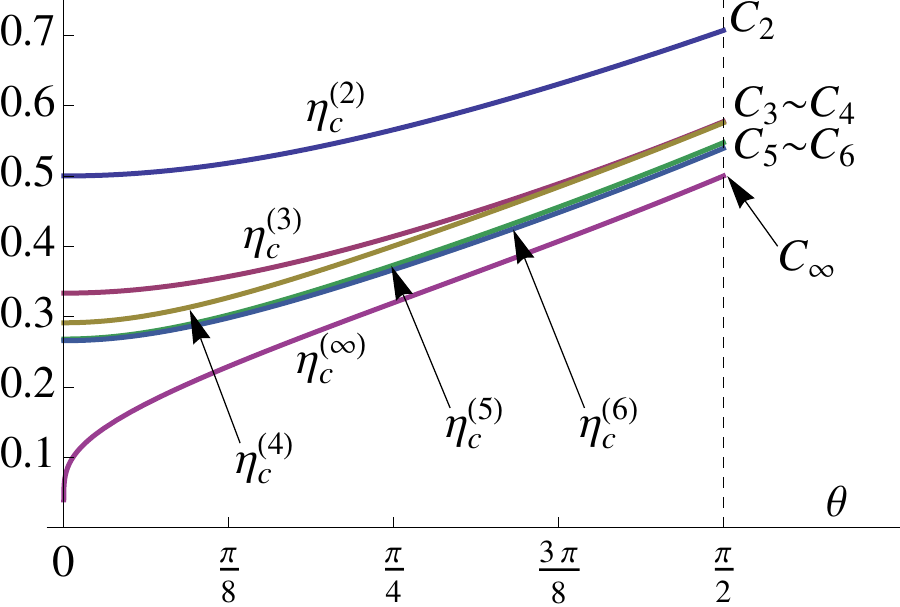}
\caption{Critical efficiency in function of the entanglement parameter $\theta$. We report the critical efficiencies for the
$n=2,\ 3,\ 4,\ 5,\ 6$ and $\infty$ setting scenario. The curves for $n=5$ is slightly higher than the $n=6$ curve.}
\label{fig:eta}
\end{figure}
The efficiency $\eta_c$ is finally minimized by the following procedure:
 choose the $\vec b_k$ such that
the eigenvalues of the operator $\frac{1}{n}\sum_k\vec b_k\vec\sigma$ are precisely $\pm C_n$
and the $-C_n$ eigenvector is the state $\ket{1}$. By this choice we get 
$\frac{1}{n}\sum^n_{k=1}\langle \vec b_k\vec \sigma\rangle_{\rho_B}=-C_n\cos\theta$.
The remaining term $\frac{1}{n}\sum^n_{k=1}\langle \vec a_k\vec \sigma\otimes \vec b_k\vec \sigma\rangle_{\rho_{AB}}$
in \eqref{etac2} can be calculated by means of \eqref{alpha}.
For instance, the obtained values for $n=2$ and $n=3$ are  $\eta^{(2)}_c(\theta)=\frac{1 - \cos\theta}{\sqrt{1 + \sin^2\theta}-\cos\theta}$
and  $\eta^{(3)}_c(\theta)=\frac{1 - \cos\theta}{\sqrt{1 + 2\sin^2\theta}-\cos\theta}$.
In the $n\rightarrow \infty$ limit we should replace the sum $ \frac{1}{n}\sum^n_{k=1}$ with the integral 
$\frac{1}{2\pi}\int^{2\pi}_0\dd \phi\int^{1}_{0}\dd \cos\theta$ by considering an infinite number of
$\vec b_k$ vector with positive $z$ component. In this case we obtained $\eta^{(\infty)}_c=1/[1 + (1 + \sec\theta)\text{arccosh}(\csc\theta)]$.

We report in figure \ref{fig:eta} the values of the critical efficiencies $\eta^{(n)}_c$ as a function of $\theta$ for the
$n=2,\ 3,\ 4,\ 5,\ 6$ and $\infty$ setting scenario. We notice that, for the maximally entangled state
 $\theta=\pi/2$, we get the
expected result of $\eta^{(n)}_c=C_n$. We define $\bar\eta^{(n)}_n$ the limit of zero entanglement, namely
$\bar\eta^{(n)}_c\equiv\lim_{\theta\rightarrow0}\eta^{(n)}_c(\theta)$.
It is worth noting that $\bar\eta^{(n)}_c$ is always lower than $C_n$ and an arbitrary low value can be obtained by
increasing the number $n$ of observables. In fact we have $\bar\eta^{(2)}_c=\frac12$, $\bar\eta^{(3)}_c=\frac13$, 
$\bar\eta^{(4)}_c\simeq0.291$, $\bar\eta^{(5)}_c\simeq0.268$,
, $\bar\eta^{(6)}_c\simeq0.266$, \dots, $\bar\eta^{(\infty)}_c=0$.


\begin{figure}[tbp]
\includegraphics[width=6cm]{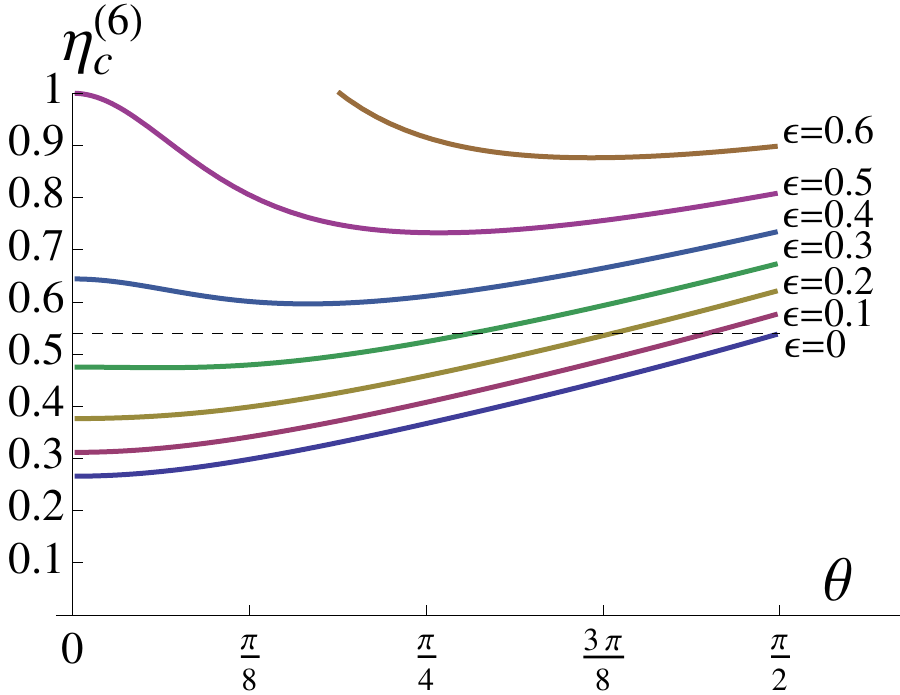}
\includegraphics[width=6cm]{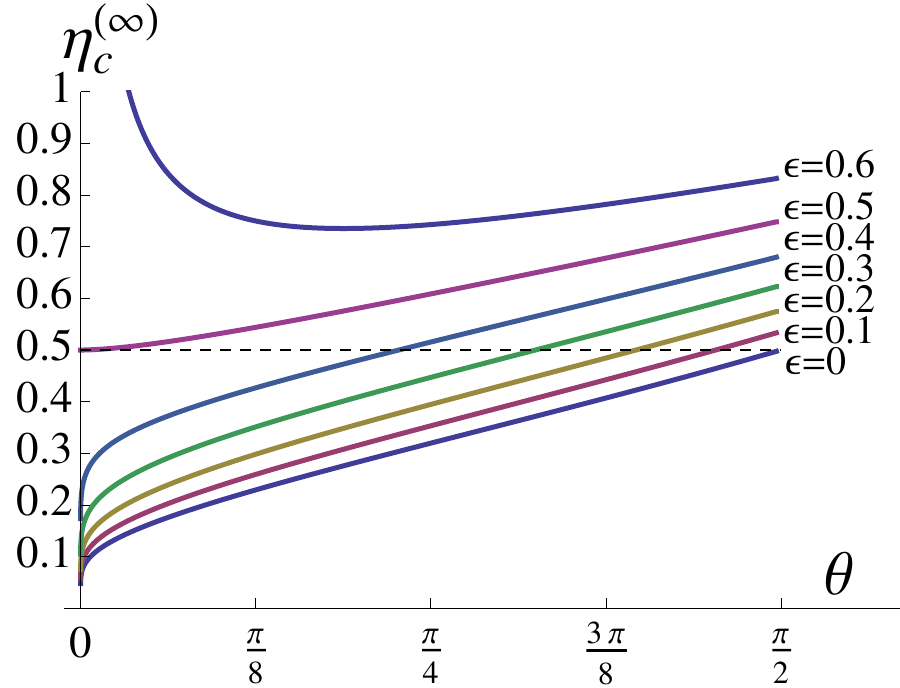}
\caption{Values of $\eta^{(noise)}_c$ in function of $\theta$ for different noise
parameter $\epsilon$ in the case of $n=6$ (top graph) and $n=\infty$ (bottom graph) measurement settings.
With dashed line we show the value of $C_6$ and $C_\infty$ respectively to compare with the efficiency required
with non-noisy maximally entangled state.
}
\label{fig:noise}
\end{figure}
We can also calculate how the critical efficiency changes if the NMES is noisy.
Here we consider a colored noise model, in which the shared state is given by
\beq\label{noise}
\begin{aligned}
\rho&=(1-\epsilon)\ket{\psi}\bra{\psi}+\epsilon\rho_{noise}
\\
\rho_{noise}&=\cos^2\frac\theta2\ket{01}\bra{01}+\sin^2\frac\theta2\ket{10}\bra{10}
\end{aligned}
\eeq
There are two main reasons to consider colored and not white (corresponding to $\rho_{noise}=\frac14\openone$) noise:
first of all, when the entangled state \eqref{psi} is experimentally generated, for example by spontaneous parametric down
conversion, the main source of imperfection
comes from the difficulty of producing $\ket{01}$ and $\ket{10}$ perfectly indistinguishable: this introduces
a decoherence precisely corresponding to our colored noise model. Moreover, the white noise
will require higher efficiency since the advantage of using NMES comes from the "polarization" of single qubit
reduced states $\rho_A$ and $\rho_B$, while white noise is completely "depolarized". On the other side, 
in the colored noise model, the reduced states $\rho_A$ and $\rho_B$ are not dependent on $\epsilon$
and the state $\rho$ is entangled for any $\epsilon>0$.
In fact, when white noise is introduced the critical efficiency is changed into
$\eta^{(white)}_c(\theta)=\left[1+\frac{\epsilon}{2(1-\epsilon)\sin^2\frac{\theta}{2}}\right]\eta_c(\theta)$ and
the limit for $\theta\rightarrow0$ is always {diverging} for any low value of the noise parameter $\epsilon$.
On the other hand, the critical efficiency $\eta^{(noise)}_c$ obtained by using the noise model \eqref{noise} can be written as
$\frac{1}{\eta^{(noise)}_c}=\frac{1-\epsilon}{\eta^{(n)}_c}+
\epsilon\frac{\frac{1}{n}\sum^n_{k=1}\langle \vec a_k\vec \sigma\otimes \vec b_k\vec \sigma\rangle_{\rho_{noise}}-C_n\cos\theta}{C_n(1-\cos\theta)}$. We show in figure \ref{fig:noise} the values of $\eta^{(noise)}_c$ in function of $\theta$ for different noise
parameter $\epsilon$ in the case of $n=6$ (top graph) and $n=\infty$ (bottom graph) measurement settings. 
It is worth noting that, even with high values of the noise, by using NMES it is possible to obtain a critical efficiency that is 
lower  than the one obtained by maximal entangled states. 
With 6 measurements and up to $35\%$ noise, NMES outperform maximally entangled states
in the required detection efficiency for a loophole free experiment.

{\it Conclusions - } In this work we showed that the inequality introduced in \cite{saun10nph} can be violated with
arbitray low efficiency by using non-maximally entangled state. This feature resembles the
property of NMES to better violate Bell inequalities in presence of detection inefficiencies. The violation of the steering
inequality is proven to be highly resistant against decoherence, the most common noise present in actual experiments.
For example, with $10\%$ noise, 
the inequality can be violated by using 6 measurements and detection efficiency larger than $31.14\%$.
Our result can have important application in quantum cryptography due to the recent connection between steering and 
cryptography~\cite{bran12pra}. This could be particular relevant
for long distance quantum communication with high losses~\cite{vill08njp}, in which
the trusted device is located at distance with respect to the entanglement source while the untrusted device is placed 
close to the source to achieve
the required efficiency needed to violate a steering inequality.

\begin{acknowledgments} 
We thanks P. Villoresi for useful discussions.
This work has been carried out within the Strategic Project QUINTET of the Department of
Information Engineering, University of Padova and the
Strategic-Research-Project QuantumFuture of the University of
Padova.
\end{acknowledgments}


%

\end{document}